\documentstyle[11pt,newpasp,twoside,epsf]{article}
\def\edcomment#1{\iffalse\marginpar{\raggedright\sl#1\/}\else\relax\fi}
\reversemarginpar

\begin{document}
\title{The Hamburg/RASS optical identification project}
 \author{F.-J. Zickgraf}
\affil{ Hamburger Sternwarte, Gojenbergsweg 112, 21029 Hamburg, Germany}
\author{D. Engels, H.-J. Hagen, D. Reimers}
\affil{ Hamburger Sternwarte, Gojenbergsweg 112, 21029 Hamburg, Germany}

\begin{abstract}
We use digitized Schmidt direct and prism  plates taken for the  northern
hemisphere Hamburg Quasar Survey (HQS) to obtain optical identifications for all 
high galactic latitude  X-ray sources  in the ROSAT 
Bright Source Catalogue (RASS-BSC) at $|b| \ge 30\deg$ and 
$\delta \ge 0\deg$. In this part of the sky the RASS-BSC 
contains about 5300 X-ray sources. Presently, identifications exist for  
$\sim4100$ RASS-BSC sources in this ga\-lac\-tic latitude range.  
\end{abstract}

\section{RASS sample}
The ROSAT Bright Source Catalogue (1RXS-B) contains a total of 
18\,811 ROSAT
all-sky survey sources with a count-rate limit of 0.05\,cts\,s$^{-1}$ in the
0.1-2.4\,keV energy range and with a typical positional accuracy of $\sim30\arcsec$
(Voges et al. 1999).
The goal of our project is to obtain optical identifications for all 1RXS-B
sources in the northern hemisphere  with $\delta \ge 0\deg$ and galactic latitudes 
$\mid b\mid \ge 30\deg$. In this part of
the sky the 1RXS-B catalogue contains 5341 sources.

\section{Method of identification}
For the identification of the  optical
counterparts we make use of the digitized objective prism and direct Schmidt 
plates of the Hamburg Quasar Survey (Hagen et al. 1995).
A description of the digital data base is given by Engels et al. (this
conference). The identification procedure is as follows:
\begin{itemize}
\item[1.] We first search on the digitized 
direct plates for all possible counterparts in the X-ray error 
circle. A minimum error circle of 40\arcsec\ is adopted.
\item[2.] Then spectra are extracted from the data base of the digitized 
spectral plates at the positions of the objects.  The limiting magnitude of the
spectral plates is $B \approx 18.5$. 
\item[3.] Using these spectra the
objects are classified interactively. Object categories used are listed in Tab.
1.
\item[4.] Finally, all possible counterparts are entered into a catalogue and 
the most likely optical counterpart is identified by a flag (cf. Fig. 1). 
\end{itemize}

\begin{figure}
\caption{Section of the catalogue.  
The most likely optical 
counterpart is flagged with a ``+'' sign. For object categories (last
column) see
Tab.~1. }
\plotone{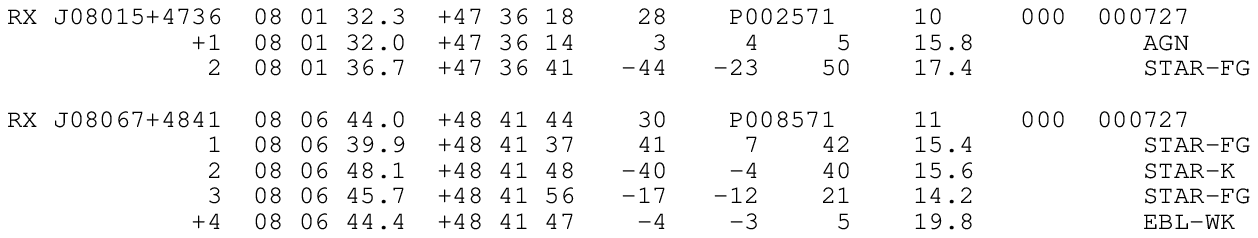}
\end{figure}

\begin{table}
\caption{Object categories used for the classification of objects 
found in the X-ray error circle.}
\begin{tabular}{ll}
\noalign{\smallskip}\tableline\noalign{\smallskip}
$category$ & $object ~properties$  \\
\noalign{\smallskip}\tableline\noalign{\smallskip}
SUBDWARF & hot subdwarf\\
W-DWARF & white dwarf\\
CV & cataclysmic variable, exhibits Balmer emission\\
STAR-BA & Balmer line absorption, point-like image\\
STAR-FG & G-band, Ca\,H\&K absorption, point-like image\\
STAR-K & G-band, Ca\,H\&K absorption,\\ 
&        redder continuum than STAR-FG\\
STAR-M & continuum very red, TiO and CaI\,4226\AA\ absorption\\
GALAXY & red continuum, no emsiison lines, extended direct image\\
BLUE GALAXY & continuum blue, extended direct image\\
AGN & continuum blue, emission lines\\
QSO & continuum very blue, emission lines, point-like image\\
EBL-WK& continuum very blue, faint point-like image\\
BLUE-WK & blue continuum, faint point-like image\\
RED-WK & red continuum, faint point-like image\\
UNIDENT & none of the above defined classes assignable\\
OVERLAP &  no classification possible due to overlapping spectra\\
SATURATE &  no classification possible due to saturated spectrum\\
NONSENSE &  spectrum of no use, e.g. plate defect, blend\\
\noalign{\smallskip}\tableline
\tableline

\end{tabular}
\label{cat}
\end{table}

\section{Status of the project}
For 4190 of the 5341 RASS-BSC sources at $\delta \ge 0\deg$ and 
$\mid b \mid \ge 30\deg$ identifications exist already. 
The current version 2.0 of the catalogue 
additionally contains  456 BSC sources with 
$20\deg < \mid b \mid < 30\deg$ (cf. Bade et al. 1998).
We are presently processing the remaining 1151 sources.
The {\em Hamburg/RASS
catalogue of optical identifications} (version 2.0) is
available on\\
\\
\centerline{http://www.hs.uni-hamburg.de/hrc.html}

\begin{table}
\caption{Current statistics of the various object classes
identified in the catalogue (version 2.0)}
\begin{tabular}{llllll}
\noalign{\smallskip}\tableline\noalign{\smallskip}
$class$ &$number$& \%~~~~~~~~~~~~&  $class$ & $number$& \% \\
\noalign{\smallskip}\tableline\noalign{\smallskip}
AGN, QSO & 1802& 43 & F/G stars& 4&0.1 \\
galaxies& 148&3.5& K stars & 129&3.1\\
galaxy clusters& 149&3.5 &M stars& 162&3.9 \\
white dwarfs& 36&0.9 & unidentified& 678&16\\
cataclysmic variables& 20&0.5& blank fields& 112&2.7\\
bright stars& 950&23 \\
\noalign{\smallskip}\tableline
\tableline

\end{tabular}
\label{stat}
\end{table}

\section{Current statistics}
The current statistics are summarized in Tab. 2. The largest group of objects 
in the catalogue are the Active Galactic Nuclei comprising 43\% of
the RASS-BSC sources, followed by the stellar counterparts which represent  31\%
of the X-ray sources. 7\% of them are K-M stars. Note that most F/G stars  
appear in the group of 
``bright stars'' comprising 23\% of stellar counterparts. Due to the brightness
the spectra of these objects are saturated and therefore a detailed 
classification is not
possible. 16\% of the sources could not be identified.

A special problem of prism surveys is the increasing rate of overlapping spectra
with decreasing galactic latitude due to the increasing surface density of stars
near the galactic plane. The 456 error circles located between $\mid
b \mid = 20\deg$ and $30\deg$ contain 1458 objects of which  135 
have overlapping spectra, i.e. 
9\%. In the error circles of the sample with $\mid
b \mid > 30\deg$  8972 objects were found with 565, i.e. 
6\%, overlapping spectra. In order to keep the rate of overlaps minimal we 
restrict the complete identification of the RASS-BSC to the region 
$\mid b \mid \ge 30\deg$.

\end{document}